\documentstyle[11pt,aaspp4]{article}
\eqsecnum

\begin{document}

\font\twelvei = cmmi10 scaled\magstep1 
       \font\teni = cmmi10 \font\seveni = cmmi7
\font\mbf = cmmib10 scaled\magstep1
       \font\mbfs = cmmib10 \font\mbfss = cmmib10 scaled 833
\font\msybf = cmbsy10 scaled\magstep1
       \font\msybfs = cmbsy10 \font\msybfss = cmbsy10 scaled 833
\textfont1 = \twelvei
       \scriptfont1 = \twelvei \scriptscriptfont1 = \teni
       \def\mit{\fam1 }
\textfont9 = \mbf
       \scriptfont9 = \mbfs \scriptscriptfont9 = \mbfss
       \def\bmit{\fam9 }
\textfont10 = \msybf
       \scriptfont10 = \msybfs \scriptscriptfont10 = \msybfss
       \def\bmsy{\fam10 }

\def\ea{{\it et al.}~}
\def\etal{{\it et al.~}}
\def\eg{{\it e.g.,~}}
\def\ie{{\it i.e.,~}}

\def \msol {\rm{M}$_\odot$}
\def \mdot {\rm{M}$_\odot$~yr$^{-1}$}
\def \kms{km~$\rm{s}^{-1}$}
\def \cc{$\rm{cm}^{-3}$}
\def \arcs{\char'175}
\def \lam{$\lambda$}
\def \micra{$\mu$m}

\def\lsim{\raise0.3ex\hbox{$<$}\kern-0.75em{\lower0.65ex\hbox{$\sim$}}} 
\def\gsim{\raise0.3ex\hbox{$>$}\kern-0.75em{\lower0.65ex\hbox{$\sim$}}} 

\title{Effects of Cooling on the Propagation of
Magnetized Jets}

\author{A. Frank\altaffilmark{1},
        D. Ryu \altaffilmark{2},
        T. W. Jones \altaffilmark{3},
         A. Noriega-Crespo\altaffilmark{4} }

\altaffiltext{1}{Department of Physics and Astronomy, University of
    Rochester, Rochester NY 14627-0171:\\
    afrank@alethea.pas.rochester.edu}
\altaffiltext{2}{Department of Astronomy \& Space Science, Chungnam National
    University, Daejeon 305-764, Korea:\\
    ryu@sirius.chungnam.ac.kr}
\altaffiltext{3}{Department of Astronomy, University of Minnesota,
    Minneapolis, MN 55455: twj@msi.umn.edu}
\altaffiltext{4}{IPAC, California Institute of Technology,
    Pasadena, CA 91125: alberto@ipac.caltech.edu}

\begin{abstract} 

We present multi-dimensional simulations of magnetized radiative jets
appropriate to Young Stellar Objects. 
  Magnetized jets subject to collisionally excited radiative
losses have not, as yet, received extensive scrutiny. The purpose of
this letter is to articulate the propagation dynamics of radiative
MHD jets in the context of the extensive jet literature.  Most
importantly, we look for morphological and kinematic diagnostics that 
may distinguish hydrodynamic protostellar jets from their magnetically
dominated cousins.

Our simulations are axisymmetric ($2{1\over2}$-D).  A toroidal ($B_\phi$)
field geometry is used.  Our models have high sonic
Mach numbers ($M_s \approx 10$), but lower fast mode Mach number ($M_f
\approx 5$). This is approximately the case for jets formed via
disk-wind or X-wind models - currently the consensus choice for launching
and collimating YSO jets. Time-dependent radiative losses are included
via a coronal cooling curve.

Our results demonstrate that the morphology and propagation
characteristics of strongly magnetized radiative jets can differ
significantly from jets with weak fields.  In particular the formation
of {\it nose-cones} via post-shock hoop stresses leads to narrow bow
shocks and enhanced bow shock speeds. 
In addition, the hoop stresses
produce strong shocks in the jet beam which constrasts with the
relatively unperturbed beam in radiative hydrodynamic jets.  
Our simulations show that
pinch modes produced by magnetic tension can strongly effect magnetized
protostellar jets.  These differences may be useful in observational
studies designed to distinguish between competing jet collimation
scenarios.

\end{abstract}

\clearpage

\section{Introduction} 

The origin of supersonic jets from Young Stellar Objects (YSOs) remains
unclear.  The current consensus holds that magnetic fields tied to
either an accretion disk or the star-disk boundary can launch {\it and}
collimate material into a jet.  These ``magneto-centrifugal" scenarios
have been explored analytically in a variety of configurations
including the popular disk-wind (\cite{Pud91}; \cite{KoRu93}) and
X-Wind (\cite{Shuea94}) models.  Numerical simulations of
magneto-centrifugal mechanisms have shown mixed but promising results.
\cite{PudOuy97} demonstrated that disk wind models can produce
well collimated jets when the feed-back on the disk is ignored.
\cite{Romanova97} have shown that magneto-centrifugal mechanisms can
launch winds; however, in their simulations the winds do not collimate
into jets.  The potential difficulties involved in turning
magneto-centrifugal winds into jets have been noted before.
\cite{LiShu96} discussed the slow (logarithmic) rate at which MHD
models produce collimation.  \cite{Ostriker97} demonstrated that
X-wind models will tend to produce wide-angle winds rather than
discrete cylindrical jets.

Along with these issues, recent numerical studies have shown that pure
hydrodynamic collimation can be surprisingly effective at producing well
collimated supersonic jets.  Frank \& Mellema (1996, 1997) and Mellema
\& Frank (1997) have demonstrated that isotropic or wide angle YSO winds
interacting with toroidal density environments readily produce oblique inward
facing wind-shocks.  These shocks can be effective at redirecting the
wind material into a jet.  If the wind from the central source is varying,
this ``shock focusing'' mechanism can, in principle, produce jets on the
observed physical scales (Mellema \& Frank 1997). Similar mechanisms
have been shown to work in other jet-producing contexts as well
(Peter \& Eichler 1996; Frank, Balick \& Livio
1996; Borkowski, Blondin \& Harrington 1997).

The variety of available collimation models begs the question of which
process actually produces protostellar jets.  Observations that can
distinguish signatures of different theoretical models are
obviously needed.  Unfortunately seeing into the collimation region is
difficult.  A number of studies indicate that collimation occurs on
scales of order $R \le 10 AU$ (Burrows \ea 1996; Ray \ea 1996), at or
below current observational limits.  In addition the many magnitudes of
extinction common for star forming systems often obscure the innermost
region where jets form.  Thus critical observations concerning the
formation of the jets will have to come from downstream of the
collimation regions; i.e. from the jets themselves.

If the collimation process is MHD dominated, then magnetic fields will remain
embedded in the jets as they propagate.  In particular both disk-wind
and X-wind models of jet collimation will produce jets with strong
toroidal fields.  This is evidenced by the fact that while the sonic
Mach numbers of the jets may be high ($M_s > 10$) the fast mode Mach
numbers will be low ($M_f \approx 3$, \cite{Camen97}). Direct observation of
magnetic fields in protostellar jets would help clarify issues surrounding 
jet origins.  Unfortunately such measurements have generally proven to be
difficult to obtain (Ray \ea 1997).   

A promising alternative is look for less direct tracers of strong fields
in YSO jets.   If jets are produced via MHD processes, dynamically
significant magnetic stresses should affect the beam and jet head as
they interact with the environment.  Thus the propagation
characteristics of protostellar jets may hold important clues to their
origins.

In this letter we present the first results of a campaign of radiative
MHD simulations of YSO jets.  The goal of our ongoing study is to search
for observable characteristics that distinguish hydrodynamic from
magnetohydrodynamic jets.  Here we present models that  
articulate significant radiative MHD effects while also making contact
with the extensive bibliography of previous numerical jet studies.
Radiative MHD jets are the next logical step in the explication of
astrophysical jet dynamics.  We have deliberately used simplified
initial conditions in our simulations. Our initial set-up,
similar to those used past studies, demonstrates new features
introduced by the interaction of radiative losses and magnetic stresses
in the context of those aspects of jet physics that are well understood.

\section{A Short History of Astrophysical Jets Simulations} 

Beginning with the work of Norman and collaborators the behavior of
axisymmetric jets ($2{1\over2}$-D) without radiative losses has been
successfully cataloged and explained (Norman 1993 and references therein).
The dynamics of the {\it bow shock} (which accelerates ambient
material), {\it jet shock} (which decelerates material in the jet
beam), and {\it cocoon} (decelerated jet gas surrounding the beam) have
been well-studied in these simulations.  One should note that these
investigations have tended to focus on so called ``light'' extra-galactic
jets where the density of the material in the jet beam ($\rho_j$) is
lower than that in the ambient medium ($\rho_a$).

The jets emanating from YSOs are, however, thought to be ``heavy'' in the
sense that the ratio $\eta = \rho_j/\rho_a \ge 1$. Another fundamental
difference between extra-galactic and YSO jets is the presence in the
latter of strong post-shock
emission from collisionally excited atomic and molecular lines.  Thus,
while extra-galactic jets can be considered ``adiabatic'', YSO jets must
be considered ``radiative''.  Beginning with the work of Blondin,
K\"onigl \& Fryxell (1990) the dynamics of radiative jets has been
explored in considerable detail (Stone \& Norman 1994; Raga 1994;
Suttner \ea ~1997).  These simulations all show that as pressure support
is lost, the bow shock/jet shock pair collapse into a thin shell. They
also revealed dynamical and thermal instabilities associated
with this shell.  

MHD simulations of non-radiative jets were first carried out by Clarke,
Norman \& Burns 1986 using a strong, purely toroidal magnetic field
${\bf B}=(0,B_\phi,0)$.  Their results showed that ``hoop'' stresses
associated with the radially directed tension force inhibit
sideways motion of shocked jet gas.  Material that
would have spilled into the cocoon is forced into the region between
the jet and bow shock, forming a ``nose-cone'' of magnetically dominated
low $\beta$ gas ($\beta = P_g/P_B = 8 \pi P_g/B^2$). Hoop stresses 
also collapse the beam near the nozzle, producing strong
internal shocks. Lind \ea (1989) performed similar calculations,
initializing their simulations with a jet in hydromagnetic equilibrium
and presenting a more complete exploration of parameter space. 
They confirmed that nose-cones form in jets with low initial
$\beta$.   K\"ossl, M\"uller \& Hillebrandt (1990ab) explored
MHD jets with a variety of initial field configurations (poloidal,
toroidal, and helical).  Cases with significant toroidal fields always
developed nose-cones.  Cases with poloidal fields developed loops with
field reversals in the cocoon susceptible to tearing mode
instabilities and reconnection.  We note studies of MHD instabilities in 
{\it just} the jet beam have also been carried out (\eg Hardee, Clarke \&
Rosen 1997)

MHD simulations of heavy YSO jets have been carried out by Todo \ea
1992.  These models did {\it not} include radiative losses.  
Their jets showed similar forms to those simulated by K\"ossl M\"uller \&
Hillebrandt 1990a, but Todo \ea 1992 also were able to identify the presence of
both slow and fast mode shocks in the jet and to present a  more
quantitative analysis of stability issues.  

Very recently, Cerqueira \ea 1997, in a study parallel to ours,
have reported the first 3-D simulations
of radiative MHD jets using an SPH code.  Those simulations show results
that are similar to what we present below.  Our work
and that of Cerqueira \ea are complimentary in that very
different methods are used.
Their study was carried out in 3-D, while ours is $2\frac{1}{2}$-D
at this point. However,our grid-based
axisymmetric simulations have a factor 10 higher resolution.
Especially in radiatively unstable structures, high numerical 
resolution is important to capture the dynamics properly.

\section{Initial Conditions} 

Our simulations evolve the equations of
ideal MHD in cylindrically symmetric coordinates ($r,z$).
Since all three components of vector fields are accounted for
the model is $2{1\over2}$-D.
The MHD code used is based on the MHD version of the Total Variation
Diminishing method.  It is an extension of the second-order
finite-difference, upwinded, conservative scheme, originally
developed by Harten (1983).
The MHD code is described in \cite{ryuj95} (1-D version),
\cite{ryujf95} (multidimensional Cartesian version), and
\cite{ryuyc95} (multidimensional cylindrical version).
The code contains routines that maintain the
${\bmsy\nabla}\cdot{\bmit B}=0$ condition at each time step.
However, since the simulations described actually contain
only the toroidal component of magnetic field in $2{1\over2}$-D,
${\bmsy\nabla}\cdot{\bmit B}=0$ is trivially satisfied although
${\bmsy\nabla}\cdot{\bmit B}=0$ is not enforced in the code.

Cooling is calculated from look-up tables for a coronal cooling curve
$\Lambda(T)$ taken from \cite{DalMc72}. Full ionization is assumed
and the cooling is applied in between hydro time-steps via an
integration of the thermal energy $E_{\rm t}$ (\cite{MelFr97}).  Tests
show the method can recover steady state radiative shocks to within
$1\%$ accuracy when the cooling region is resolved.  A ``floor" on the
temperature is set at $T = 10^4 K$.  The code has been extensively tested in 
1.5 dimensions (Franklin, Noriega-Crespo \& Frank 1997) with and without 
cooling and against Uchida et al. simulations (1992) with satisfactory results.

We have performed simulations that compare the evolution of
4 cases: a non-radiative weak field jet; a non-radiative strong field
jet; a radiative weak field jet; a radiative strong field jet.  Since
the first three cases have been studied before, our goal was to confirm
that the code recovers features seen in previous investigations available
in the literature and to extend the sequence into the radiative,
strong-field regime.

In each simulation the jet was driven into the computational domain
($128 \times 1024$) as a fully collimated supersonic/super-fast mode
beam ${\bf v} = (0,0,v_z)$.  The properties of the jet common to all
the simulations were as follows ($j$=jet, $a$=ambient medium): $\eta =
n_j/n_a = 1.5$; $n_j = 90 ~cm^{-3}$; $T_a = 1.5\times 10^4 ~K$; $v_z =
100 ~km s^{-1}$.  The initial gas pressure in the jet $P_j$,
and hence $T_j$ are varied radially to obtain hydromagnetic equilibrium.  
Thus the sonic Mach number
$\bar{M}_s$ must be defined as a radial average of $v_j/(\sqrt{\rho_j/\gamma
P_j(r)}$.  In all our simulations $\bar{M}_s \approx 10$ The
computational domain spanned $(R,Z) = (8.5 \times 10^{16} ~cm, 6.8
\times 10^{17} ~cm)$ with a jet radius $R_j = 2 \times 10^{16} ~cm$ or
30 grid cells.

In all the simulations a magnetic field was imposed in the {\it jet
only}.  The field was purely toroidal.  This simplification can be
justified on theoretical grounds (Camenzind 1997; \cite{PudOuy97}),
since most disk-wind models rely on a dominant toroidal field to
produce tightly collimated jets. Hydromagnetic equilibrium between
gas and magnetic pressure in the jet was
imposed as an initial condition (Priest 1983, Begelman 1997)
After choosing a form for $B_\phi (r)$ the equilibrium condition is
solved for the initial radial
gas pressure distribution $P_j(r)$.  In
our simulations we used the same $B_\phi (r)$ and $P_j(r)$ as Lind \ea
(1989). As with the sonic Mach number the Alfve\'nic Mach number $M_a$ 
for the jet is a radial average of $B_\phi(r)/\sqrt{4\pi \rho_j}$.  
The fast mode, sonic and Alf\'venic Mach numbers are related by
$M_f^{-2} = M_s^{-2} + M_a^{-2}$. For our strong field simulations the
average initial value of the field is $\approx 100$ $\mu G$ and an average 
plasma beta $\approx .7$. 
It should be noted that this configuration, known as a
Z-pinch in the plasma physics community, is almost always unstable to
both pinch and kink modes. Since our simulations are axisymmetric we
are unable to track kink modes (but, see Cerqueira \ea 1997; Todo \ea 1993).
The pinch modes are quite important, however, and, as Begelman 1997 has
shown, the beam will be unstable to these instabilities when
\begin{equation}
{d \ln B \over d \ln r} > {\gamma \beta - 2 \over \gamma \beta + 2},
\end{equation}
where $\gamma$ is the ratio of specific heats. This condition is
satisfied for the Lind \ea 1989 initial configurations.  We note that since
these simulations include radiative cooling $P(r)$ quickly flattens
out as the jet propagates. 

\section{Results}
We first present an equation to estimate the speed of the jet head, $v_h$ (the
bow shock) that accounts for magnetic (and gas) pressure effects.  We
define $P_t = P_B + P_g$ and $\alpha = (R_j/R_h)^2$ where $R_h$ is 
effective jet
head radius over which the ram pressure, $\rho_a v^2_h$ is applied. 
>From 1-D momentum balance, ignoring the ambient pressure, we find
\begin{equation}
v_h = v_{ho} { {1~-~\sqrt{\frac{1}{\eta\alpha}~-~\frac{P_t}{\rho_j v^2_j}
(1~-~\frac{1}{\eta\alpha})}}\over{1~-~\frac{1}{\sqrt{\eta\alpha}}}},
\label{BigE}
\end{equation}
where $v_{ho} = \frac{v_j}{1~+~1/\sqrt{\eta\alpha}}$ is the familiar
expression for the speed of  a ``cold'', pressure-less jet (\eg Dal Pino
\& Benz 1993), and the remaining terms describe an ``enhancement 
factor"  for the head speed due to finite jet pressure. This factor in
equation \ref{BigE} is {\it always greater than or equal to unity}, 
as we should expect. 
For the limiting case $\eta\alpha = 1$ equation \ref{BigE} gives
$v_h = \frac{1}{2} v_j (1~+\frac{P_t}{\rho_jv^2_j})$.
In Fig 1 and 2 we present the results of our simulations with $\eta = 1.5$.
We will use equation 4.1 to interpret the results presented below.

\noindent {\bf A) Non-radiative Weak Field Jet}: Here the magnetic
field is set quite low so that Alfv\'enic Mach number is $M_a \approx
10^4$.  Fig 1 shows clearly the well-known bow shock/jet shock
configuration at the head of the jet (Norman \ea 1993). The scale of
the bow shock is relatively large because of the post-shock
thermal pressure.  Note however that the speed of the jet ($100$ \kms)
is relatively low compared with extragalactic jets.  Thus the post
{\it jet} shock temperature in this simulation is
low ($T \approx 10^5$). The pressure in the cocoon is therefore also small compared with
extragalatic jet simulations and only weak internal waves are forced
into the the jet beam.  Blondin, Fryxell \& K\"onigl 1990 found a similar
result in their comparisons of non-radiative and radiative YSO jet
simulations.

\noindent {\bf B) Non-radiative Strong Field Jet}: Here a stronger field is used
so that the Alfv\'enic Mach number is $\bar{M}_a \approx 7$ ($\bar{M}_f
\approx 5$).  In this simulation the structure of the jet head has
changed dramatically due to dynamical influences of the field.  Note the
large separation of the jet and bow shocks.  As has been described in
other MHD jet studies (Clarke \ea 1986; Lind \ea 1989; K\"ossl \ea
1990ab) such ``nose-cone'' structures occur because of the pinching
effect of the magnetic hoop stresses. Post jet-shock material, which
would otherwise back flow into the cocoon, is forced to remain between
the two shocks.  The bow shock is accelerated forward producing higher
propagation speeds (note the times at which the images where taken).  In
previous studies the acceleration was attributed to the magnetic
pressure in the nose cone (K\"ossl \ea 1990b). We demonstrate below that
the increased speed can be better attributed in this case to a cross section effect
produced by magnetic pinch forces. Note the strong convergence occurring
just beyond the jet nozzle and the reflected shocks in the beam
downstream.  This occurs due to the pinch instability described in 
section 3 and has been seen in all MHD jet
simulations with strong toroidal fields. We note also that these results
look quite similar to the those presented by K\"ossl \ea 1990b (see
their figure 10a).

\noindent {\bf C) Radiative Weak Field Jet}: The magnetic field in
this case is the same as case A. Again the standard bow
shock/jet shock configuration is apparent.  Maps of temperature show
that the post-bow/jet shock region is nearly isothermal. The loss of
post-shock thermal energy reduces the transverse width of the bow shock. 
Densities behind the shocks become high with compression ratios of 
$\approx 30$ (compared to the value $\approx 4$ obtained in the
non-radiative case).  Animations of this model show that the structure
at the jet head is highly time-dependent with the region between
the two shocks becoming quite thin at times.  At late times in the
simulation the jet head undergoes the non-linear thin shell instability
(Vishniac 1994), which also has been observed in other radiative jet
simulations (Blondin, Fryxell \& K\"onigl 1990; Stone \& Norman 1994).
Note again the lack of structure in the jet beam. In both cases A)
and C) the jet heads propagate at a velocity $v_{h} \approx 53$ km
which compares well with $v_{ho} = 55$ predicted by
equation \ref{BigE} for a cold jet ($v_{ho}$) with  $\alpha \approx 1$.

\noindent {\bf D) Radiative Strong Field Jet}: In this simulation the
field is the same as for the non-radiative, strong field jet (case B)
and the structure of the jet head is similar to what obtains in that
simulation.  A narrow bow shock appears some distance ahead of a high
density region associated with the jet shock.  Note that the loss of
thermal energy behind the shocks has reduced the scale of the bow shock
$R_h$ compared with its non-radiative twin. Fig 2 shows the evolution
of this simulation at five different times.

For this case the initial collapse of the beam just beyond the nozzle
produces a series of strong shocks and reflections in the jet as it
propagates down the length of the grid. Fig 2 shows the periodic
density enhancements in the beam formed from these reflections.  If
such features are not a consequence of the imposed axi-symmetry they
may have important consequences for the emission characteristics of
real YSO jets. Figs 1 and 2 show that a jet shock forms close to the
bow shock.  The reduced width of the jet shock/bow shock pair is
similar to what is seen in the evolution of the weak field radiative
jet (Blondin, Fryxell \& K\"onigl 1990).

The most important conclusion to be reached from these simulations
comes from comparison of the third and fourth panels of Fig 1.  For
radiative jets, the weak and strong field cases look dramatically
different from each other in terms of the morphology of the jet head
and beam.  In the weak field case there are no shocks in the beam.  The
strong field case shows multiple shock reflections (Fig 2). The head of
the weak field jet is quite ``blunt'' compared with the strongly
tapered strong field jet. The average propagation speed for the head of
the strong field jet is $v_{h} \approx 70$ \kms, a 30\% increase over
the weak field case, even though both simulations have the same
jet/ambient density ratio, $\eta$.  The combination of higher shock
speeds and strong pinch forces in the radiative strong field jet
produces a compression ratio in the head almost twice as large as in
the radiative weak field case. If these results are born out in more
detailed studies, particularly those in 3-D, then indirect diagnostics
of the presence of dynamically strong fields in jets should exist.

As noted, the increased speed of MHD jet heads has sometimes been
attributed to magnetic pressure in the nose-cone. Equation \ref{BigE},
however, demonstrates that the increased momentum provided by
finite pressure alone cannot account for the enhanced head speed in the MHD
cases we have simulated.  Using  $\eta = 1.5$, $\bar{M_s}
 \approx 10$, $\bar{M_a} \approx 7$ appropriate to our simulations,
this relation gives  $v_{h} = 56 ~km s^{-1}$, assuming the same
head-scale factor $\alpha = 1$ that was used successfully to estimate
the jet head speeds for our non-MHD simulations. In fact, the finite
pressures nominally add only about 2\% to the jet head speeds according
to equation \ref{BigE}.  On the other hand, it is apparent that the
tapered shape of the nose cone formed by magnetic hoop stresses has
streamlined the flow around the jet head. That effectively increases
the geometry factor, $\alpha$, which can significantly enhance the jet
head speed, $v_h$.  From the simulation data we estimate in case D that
the radius of the jet head is $R_h \approx 8\times10^{15}~cm$, leading
to $\sqrt{\alpha} \sim ~ 2.5$. Inserting this into equation \ref{BigE}
gives a jet head speed of  $77 ~ km  s^{-1}$, closer to the $70 ~km
s^{-1}$ estimated from the simulation itself.

\section{Conclusion and Discussion} 

The results of these simulations demonstrate that propagation-based 
diagnostics for radiative MHD jets may exist. 
The strong effect of magnetic pinches
in these axisymmetric calculations change both the structure of the jet
head and the beam.  The increased velocity of the bow shock is also a
distinctive feature of MHD jets. These morphological and kinematic
characteristics would alter the observed emission properties in a
real YSO jet. The strong shocks in the beam would produce increased
excitation of both atomic and molecular lines.  The increased speed of
the jet head will alter both the degree of ionization and excitation.
This study is too preliminary, however, to provide
observers with a definitive accounting of the differences between real
radiative hydrodynamic and MHD jets.  We leave this task to future studies
(Frank \ea 1998).

The most obvious deficit in our models is the imposed axisymmetry.
Cerqueira \ea 1997 have recently reported 3-D SPH calculations which
also show that strong toroidal field components dramatically alter the
morphology of the jets, consistent with our results.  Their numerical
resolution was too low, however, to see the detailed structure of
shocks in both the beam and the jet head, which are both clearly
captured in our simulations. Since these kinds of structures are not
seen in observations, Cerqueira \ea 1997 (who did observe the narrowing of
the jet head due to pinch forces) concluded that real YSO jets can
not have significant toroidal fields.  If future high resolution
grid-based and SPH based codes continue to find such effects, that may
pose a serious challenge to MHD jet models that rely on collimation via
hoop stresses.

It is also worth noting the role of reconnection.  The initial
conditions used in our models do not allow for field reversals to
occur.  If a poloidal ($B_z$) component exists, then field lines
embedded in the beam will be decelerated upon passage through the jet
shock.  If the fields and cooling are not strong enough to inhibit the
formation of a cocoon, these lines will be carried backwards forming
field reversals (Frank \ea 1997; K\"ossl, M\"uller \& Hillebrandt
1990b).  Such a topology is unstable to resistive tearing mode
instabilities and magnetic reconnection.  The presence of reconnection
in jets could have important consequences for the interpretation of
shock emission diagnostics (Hartigan, Morse \& Raymond 1993).
Reconnection would provide an alternate means for converting kinetic
energy into a thermal energy (the field acts as a catalyst, Jones \ea
1997) which is then channeled into collisionally
excited emission.

\acknowledgments

We wish to thank Ralph Pudritz, Jack Thomas and Guy Delemarter for the
very useful and enlightening discussions on this topic.  Support for
this work was provided at the University of Rochester by NSF grant
AST-9702484 and the Laboratory for Laser Energetics, and at the
University of Minnesota by NSF grants AST93-18959, INT95-11654 and
AST96-16964, by NASA grant NAG5-5055 and the University of Minnesota
Supercomputing Institute.  The work by DR was supported in part by
KOSEF through the 1997 Korea-US Cooperative Science Program.

\clearpage

\begin{center}
{\bf FIGURE CAPTIONS}
\end{center}
\begin{description}

\item[Fig.~1] 
{Comparison of 4 YSO Jet simulations. Grey-scale
Log density.  A: Weak Field Non-Radiative (t = 1,971 y). B: Strong
Field Non-Radiative (t = 1,460 y). C: Weak Field Radiative (t = 1,971
y).  D: Strong Field Radiative (t = 1,460 y). Note that compression
ratios behind  bow/jet shocks in the radiative models are higher by an
order of magnitude compared with the non-radiative cases. 
Each frame has been auto-scaled to its min/max resulting in different
greyscale ranges.}

\item[Fig.~2] 
{Propagation of the radiative strong field model at 4 times
(From top to bottom t = 698, 1398, 2095, 2793 y).
Each frame has been auto-scaled to its min/max resulting in different
greyscale ranges.  The min/max values in the frames are approximately
$5~cm^{-3}$ and $7000~cm^{-3}$.}

\end{description}

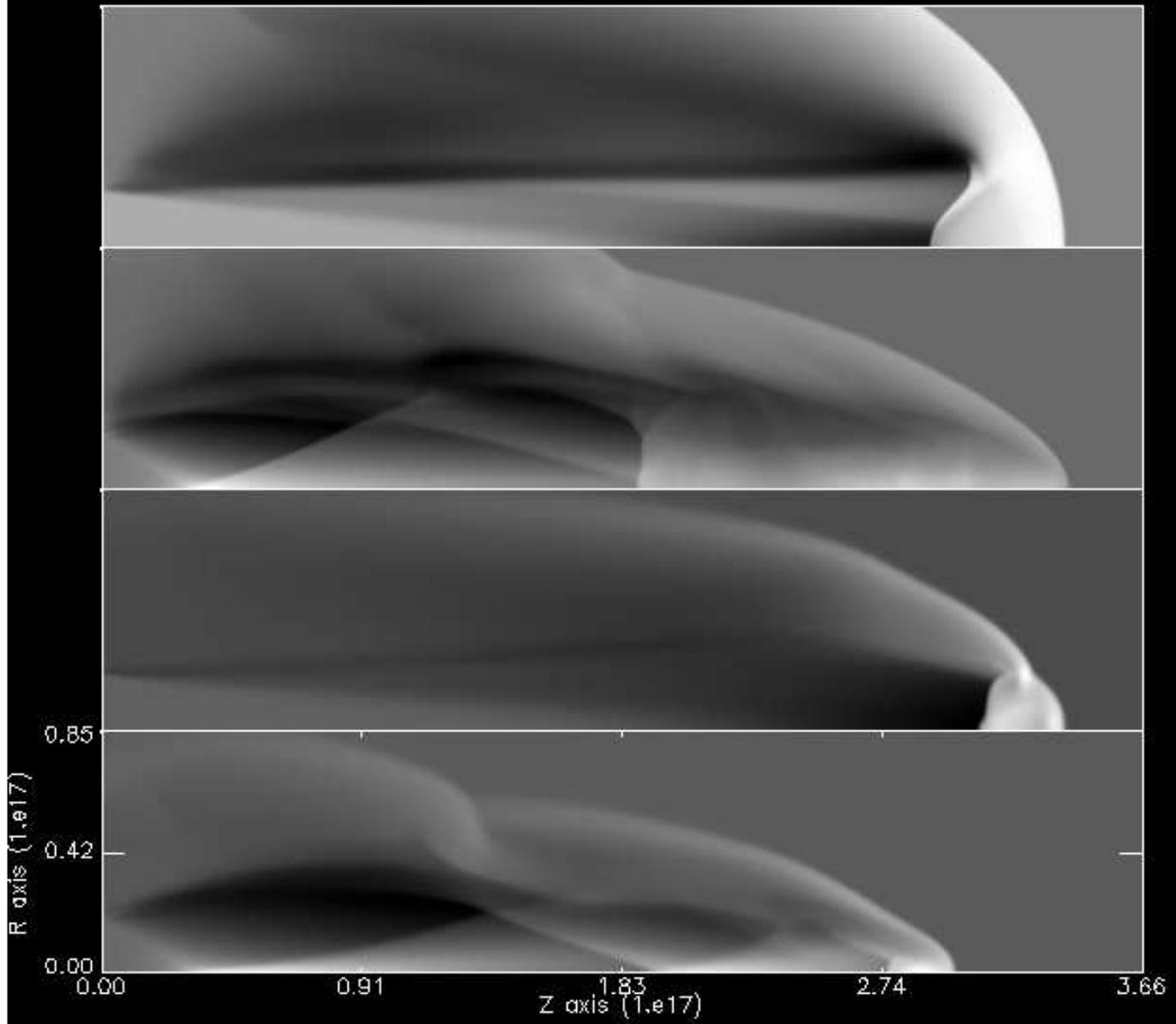
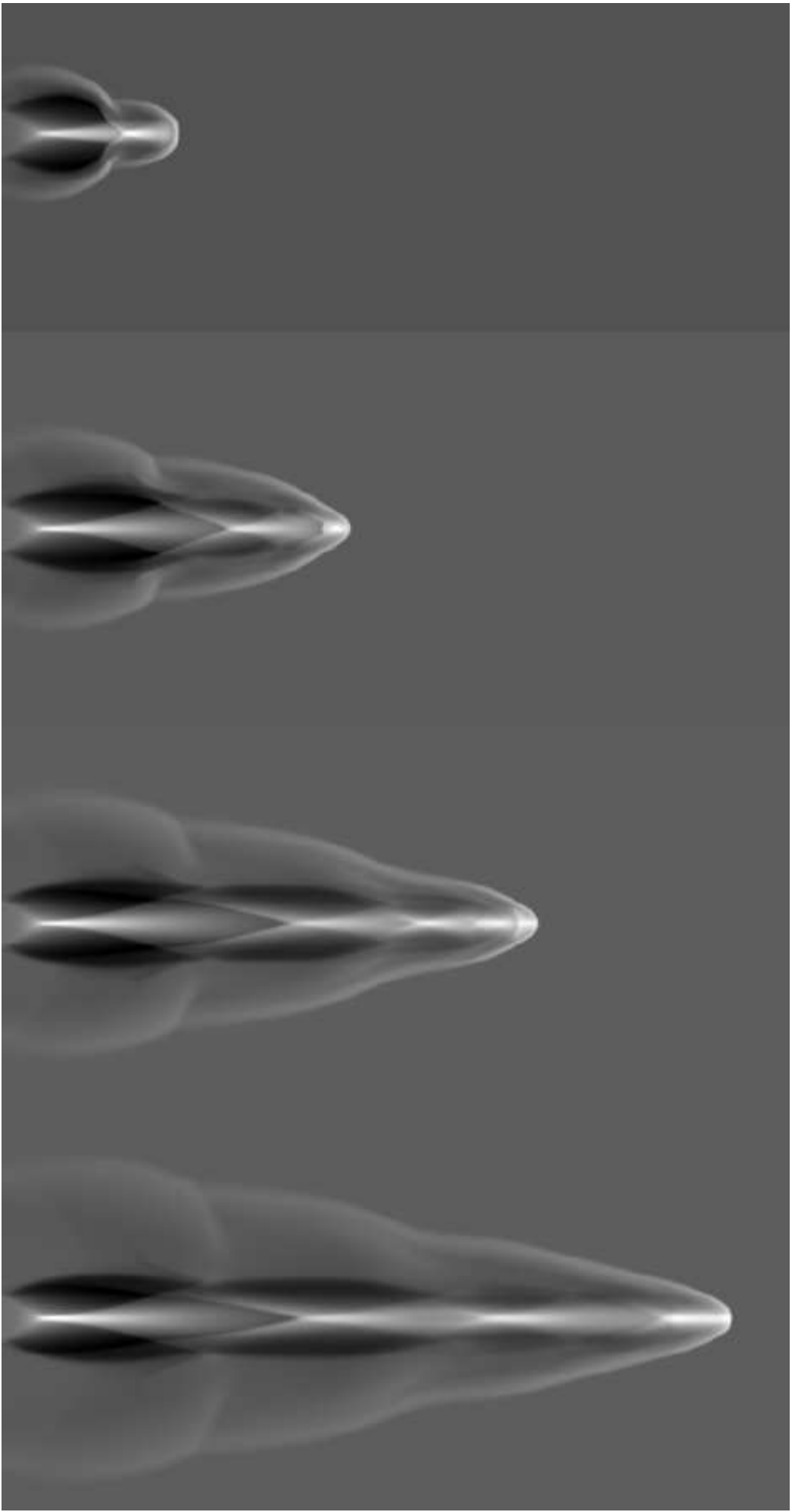
\end{document}